\documentclass[a4paper,12pt]{article}
\include{headers}
\include{math_header}
\usepackage{bm}
\usepackage{mathptmx}       
\usepackage{helvet}         
\usepackage{courier}        
\usepackage{makeidx}         
\usepackage{graphicx}        
\usepackage{multirow}        
\usepackage{amsfonts}
\usepackage{amsmath, amsthm, amssymb}
\usepackage{natbib}
\usepackage{color}
\usepackage{bm}
\usepackage{fullpage}
\usepackage{slashbox}
\usepackage{rotating}
\usepackage[english]{babel}
\newcommand{\Yb}{{\mathbf Y}}
\newcommand{\Db}{{\mathbf D}}
\newcommand{\Wb}{{\mathbf W}}
\newcommand{\Zb}{{\mathbf Z}}
\newcommand{\Gb}{{\mathbf G}}
\newcommand{\Mb}{{\mathbf M}}
\newcommand{\Sb}{{\mathbf S}}

\newcommand{\Ib}{{\mathbf I}}
\newcommand{\iw}{{\bm \Psi}}
\newcommand{\Xb}{{\mathbf X}}

\newcommand{\qb}{{\mathbf q}}
\newcommand{\Vb}{{\mathbf V}}
\newcommand{\ub}{{\mathbf u}}
\newcommand{\db}{{\mathbf d}}
\newcommand{\yb}{{\mathbf y}}
\newcommand{\epb}{{\bm \epsilon}}
\newcommand{\upb}{{\bm \Upsilon}}
\newcommand{\eb}{\mathbf{e}}
\newcommand{\bb}{{\bm \beta}}
\newcommand{\ssb}{{\bm \sigma}}
\newcommand{\xb}{{\mathbf x}}

\begin{document}


\title{Estimators of Binary Spatial Autoregressive Models:\\ A Monte Carlo Study}

\author{Raffaella Calabrese\\University of Milano-Bicocca\\raffaella.calabrese1@unimib.it \and
Johan A. Elkink\footnote{Corresponding author.}\\University College Dublin\\jos.elkink@ucd.ie}

\maketitle

\begin{abstract}

The goal of this paper is to provide a cohesive
description and a critical comparison of the main estimators proposed in the
literature for spatial binary choice models. The properties of such
estimators are investigated using a theoretical and simulation
study. To the authors' knowledge, this is the first paper that
provides a comprehensive Monte Carlo study of the estimators'
properties. This simulation study shows that the Gibbs estimator
\citep{LeSage2000} performs best for low spatial autocorrelation,
while the Recursive Importance Sampler \citep{BeronVijverberg2004} performs
best for high spatial autocorrelation. The same results are
obtained by increasing the sample size.  Finally, the linearized
General Method of Moments estimator \citep{KlierMcMillen2008} is the
fastest algorithm that provides accurate estimates for low spatial
autocorrelation and large sample size.

\end{abstract}

\section{Introduction}

In applied work in economics and political science, there is
increased attention to the importance of spatial or network
interdependence between observations. Not only does this violate the
assumption of independence underlying many econometric methodologies
for cross-sectional data, there is also growing interest in
estimating the strength of the interdependence itself. While the econometric
literature on linear regression models with spatial interdependence is well
established, in particular since the publication of \citet{Anselin1988}'s
seminal work, the literature on regression models with binary dependent
variables and spatial interdependence is still relatively limited.

Many applications with such models can be considered --  including the
contagion of currency crises \citep{Novo2003}, firm-level decision-making on
locations \citep{AutantBernard2006}, ecological studies of spatial
distributions of plants \citep{Collingham_etal2000}, studies in policy
diffusions of flat taxes \citep{BaturoGray2009}, anti-smoking laws
\citep{ShipanVolden2006} or pension privatization \citep{Weyland2007} -- across
academic disciplines such as economics, political science, sociology, ecology,
planning, or even neurology.

Proximity in this context can be interpreted in a broad manner. Whether one
defines proximity in a physical or in a cultural or interaction sense, or in a
manner that encompasses large distances or the entire space (all units affect
all other units), the estimation challenges discussed in this article still
hold.\footnote{The complications with the interpretations of the observed
effects increase, however, since factors that affect the similarity are also
likely to have an effect on the linkages between the units.  See
\citet{ShaliziThomas2010} for a discussion of the inherent confounding of
homophily and contagion mechanisms.} The conclusions can thus be directly
applied to social network analysis as well as spatial econometric analysis.
\citet[255]{Anselin2002} refers to this perspective as the {\it object view} or
this type of data as {\it lattice data}. The alternative, a {\it geostatistics}
perspective, where we observe only specific monitoring sites and space is seen
as a continuous space or a point pattern \citep{Bivand1998}, leads to an
entirely different econometric framework and will not be discussed in his
article.

Spatial econometric models raise new difficulties that cannot be dealt with by
standard econometric models.  Estimation problems arise due to the dependence
across observations, in that we must adjust the estimation procedures for the
loss of information associated with dependent observations. Indeed, in the
presence of spatial dependence, standard logit or probit estimation procedures,
which assume independence, result in inconsistent and inefficient estimates
\citep{McMillen1992}. In particular, \citet{McMillen1992} notes that both the
spatially dependent error model and the spatial lag model imply heteroskedastic
disturbances, which cause the parameter estimates to be inconsistent. For these
reasons econometricians began to pay more attention to spatial dependence
problems in the last two decades and some important advances have been made in
both theoretical and empirical studies \citep{AnselinFloraxRey2004}.

The aim of this article is to compare the main estimators proposed in the
literature for estimating the spatial autocorrelation parameter in binary
choice models. On the one hand, this goal is achieved by analysing the
theoretical characteristics of the main estimators for spatial models for
binary response data.  This topic has been in part developed by
\citet{Fleming2004} but we consider also the recent literature. Moreover, our
paper is focused only on binary choice models, instead \citet{Fleming2004} has
considered discrete choice models. On the other hand, the most innovative
aspect of this work is the comparison of the above-mentioned estimators by
Monte Carlo simulations.  To our knowledge, this is the first work that
performs Monte Carlo simulations on the main estimators of the spatial
autocorrelation parameter for binary response data. The importance and the
necessity of this analysis is strongly suggested by \citet{Fleming2004}.

Currently, the most used methodologies available to estimating spatial
regression models are five.  \citet{McMillen1992} proposes an EM algorithm
based estimation procedure. In particular, \citet{McMillen1992} replaces the
latent continuous variable with its expected value and then applies the maximum
likelihood method \citep{Ord1975}. Similarly to \citet{McMillen1992},
\citet{LeSage2000} also replaces the latent continuous variable with its
expected value, solving thereafter a spatial continuous model using the Gibbs
sampling approach.  Following the work of \citet{Vijverberg1997} on the
simulation from a multivariate normal distribution, \citet{BeronVijverberg2004}
suggests to apply the recursive importance sampling (RIS) to the maximum
likelihood method, since the likelihood function is a multivariate normal
distribution.  \citet{PinkseSlade1998} develop a model based on the generalised
method of moments (GMM).  \citet{KlierMcMillen2008} linearize
\citet{PinkseSlade1998}'s model around a convenient starting point.

The present paper is organized as follows. The next section reviews the widely
used specifications of the binary choice models with spatial dependence. In
section 3 we analyse and compare the main methodologies proposed in the
literature to estimate the spatial autocorrelation parameter in binary response
models. In section 4 we compare the properties of these estimators by Monte
Carlo simulations.  The last section concludes.

\section{Spatial binary choice models}

A widely used representation of a regression model for
an observed dichotomous response $Y_{i}$ is the latent
response model \citep[p.180]{Verbeek2008} with
dependent variable the continuous variable $Y_{i}^*$,
whereby
\begin{equation}\label{eq:observed variable}Y_{i} =\left\{
      \begin{array}{ll}
      1, & \hbox{$Y_{i}^* > 0$} \\
      0, & \hbox{otherwise,}
    \end{array}
   \right.\\
  \end{equation}
with $i=1,2,\dots,n$. A linear model is specified for
this latent response, so the model specification is
\begin{eqnarray}\label{eq:basemodel}
    \Yb^* &=&\rho \Wb \Yb^* + \Xb\bb+\db\\
    \db&=&\lambda\Sb\db+\epb, \nonumber
\end{eqnarray}
where $\Yb^*$ is a continuous random vector, $\Xb$ represents
an $n\times k$ matrix of explanatory variables, the error
term $\epb$ can follow a multivariate normal distribution in
a probit model or a multivariate logistic distribution in a
logit model.  $\Wb$ and $\Sb$ are spatial lag and spatial
error weights matrices, respectively, $\rho$ and $\lambda$
the associated scalar parameters. We highlight that only the
latent variable can be used for the spatial lag, since both
the models $\Yb^* = \rho \Wb \Yb + \Xb \bb + \epb$ and $\Yb =
\rho \Wb \Yb + \Xb \bb + \epb$ are infeasible
\citep{Anselin2002, BeronVijverberg2004, KlierMcMillen2008}.

Evidence of the absence of a consolidated literature is
given by the different denominations of the models --
we follow \citet{LeSage2000}'s notation. From the
general model (\ref{eq:basemodel}) two models are
derived.  Setting $\Sb=\mathbf{0}$ produces a spatial
lag model, which we will refer to as the Binary
Spatial AutoRegressive model (BSAR):
\begin{equation}
\Yb^*
= (I-\rho\Wb)^{-1}(\Xb\bb+\epb)
= (\Ib-\rho\Wb)^{-1}\Xb\bb+\eb,\label{eq:SAL}
\end{equation}
where
\begin{equation}\label{eq:generalized residuals}
\eb=(\Ib-\rho\Wb)^{-1}\epb.
\end{equation}
Letting  $\Wb=\mathbf{0}$ results in a regression model
with spatial autocorrelation in the disturbances, a
spatial error model which we label the Binary Spatial
Error Model (BSEM):
\begin{equation*}
\Yb^*
= \Xb\bb+(\Ib-\lambda\Sb)^{-1}\epb
= \Xb\bb+\ub,
\end{equation*}
where
\begin{equation*}
\ub=(\Ib-\lambda\Sb)^{-1}\epb.
\end{equation*}

The two models are based on different assumptions about the causes of the
spatial dependence.\footnote{For a clear interpretation of the spatial lag and
spatial error models, see \citet{Case1992}.} The spatial lag relates to an
explicit spillover effect where one agent copies behavior from neighboring
agents. It also relates to a theoretical model where the behavior is dependent
on shared resources between different agents.  The spatial error model concerns
different causal relationships. For example, a typical issue that leads to
spatial correlation in the errors is a mismatch between the spatial delineation
of the measurement and the empirical presence of the variable of interest. For
example, when studying the presence of a particular natural resource in
particular countries, the geographical zones in which this resource is present
do not usually match exactly with the country borders. A measurement of the
    presence of these resources in countries is thus necessarily spatially
    correlated, but as a nuisance rather than in a theoretically interesting
    sense.  Another common cause of spatial autocorrelation in the errors is an
    omitted variable that is itself spatially correlated. In terms of estimation,
    the two types of autocorrelation are often difficult to distinguish
    \citep[184-185]{Brueckner2003}.  The different theoretical mechanisms are of
    course not mutually exclusive and a spatial model that incorporates both a
    spatial lag and spatial residuals is perfectly reasonable.

In this paper we are primarily interested in estimating diffusion effects, and
thus our focus is on the estimation of the spatial autocorrelation parameter
$\rho$. For this reason, in this work we only analyse models with spatial lags
(BSAR) and leave spatial errors (BSEM) aside.

The contiguity or weight matrix $\Wb$ is defined by \begin{equation*} w_{ij} =
\begin{cases} 1 & \text{if the $i$-th and $j$-th observations are contiguous;}
\\ 0 & \text{if $i=j$ or the $i$-th and $j$-th observations are not
continguous,} \end{cases} \end{equation*} so it is a square matrix of order $n$
and its main diagonal elements equal to zero.  Contiguity can refer to
geographical and alternative vicinity. The use of the weight matrix $\Wb$
implies that the spatial sites form a countable lattice \citep{Lee2004}, but
part of the literature considers a continuous spatial index \citep{Conley1999}.
Because of the potential of heteroscedasticity due to the variation in the
number of neighbors for different observations, $\Wb$ is commonly normalized as
follows $w_{ij}/(\sum_j w_{ij})$ for $i,j=1,2,\dots,n$. This means that the
normalized matrix $\Wb$ is generally asymmetric, while the original weight
matrix $\Wb$ is often symmetric.\footnote{\citet{Novo2003} considers an
asymmetric non-normalized $\Wb$.} Although this is the common approach, there
are various other ways of defining and normalizing $\Wb$
\citep{Tiefelsdorf2000,Anselin2002}. Since the aim of this paper is the
comparison of the main methodologies to estimate the spatial autocorrelation
parameter $\rho$, we consider for all of them the normalized matrix
$\Wb$.\footnote{When the normalized contiguity matrix $\Wb$ is considered, to
ensure the invertibility of the matrix $(I-\rho\Wb)$ in the maximum likelihood
method, \citet{Anselin1982} proves that $1/\omega_{min}<\rho<1$ where
$\omega_{min}$ is the minimum eigenvalue of the contiguity matrix $\Wb$.}

For binary dependent variables, the most used models are the logistic and the
probit models \citep{McCullaghNelder1989}.  In the next section we analyse and
compare the main estimators of the autocorrelation parameter in both spatial
probit \citep{BeronVijverberg2004,LeSage2000,McMillen1992} and logit
\citep{KlierMcMillen2008} models.\footnote{There are a number of related
estimators that, for various reasons, will not be included in the discussion
and Monte Carlo analyses.  These estimators are related, but make assumptions
about the data that are beyond the scope of this paper.  For the spatial random
effects probit \citep{Case1991, Case1992}, when $\bf{W}$ is constrained to be
block-diagonal, in other words, when the focus is on membership of a particular
geographic region or cluster of units rather than some kind of proximity
measure, the spatial model can be substantially simplified \citep{Case1991,
Case1992}. The logistic auto-logistic \citep{GumpertzGrahamRistaino1997,
BeeEspa2008} applies to data on a regular grid, which is not applicable in the
type of diffusion studies we have in mind in this paper.  \citet{Dubin1995}'s
spatial logit model is a straightforward diffusion model that avoids most
complications of spatial models by using the temporally lagged, realized
dependent variable to create the spatial lag.  \citet{McMillen1992,
McMillen1995}'s heteroscedastic probit using weighted least squares applies to
the spatial error model, but not the spatial autoregressive model we discuss in
this paper.}

\section{Estimators for binary spatial autoregressive models}

\subsection{Expectation-Maximization algorithm}

The Expectation-Maximization (EM) algorithm is designed for cases where the
data is incomplete, for example due to missing values
\citep{DempsterLairdRubin1977}.  Since the probit model can be viewed as a
latent response model, and this latent variable is similarly unobserved,
\citet{McMillen1992} proposes to apply the EM algorithm to the probit model
with spatially lagged dependent variables and spatial error autocorrelation. In
particular, the latent unobserved observations $y_i^*$ are replaced by
estimated values. Given estimates of the values $y_i^*$, the EM algorithm
proceeds to estimate the other parameters in the model using the maximum
likelihood method.

In the EM algorithm the assumption of homogeneity for the
disturbances $\epb$ is introduced. This means that the error term
$\epb$ can follow the $n$-dimensional multivariate normal
distribution $\epb\sim N_n(\mathbf{0},\sigma^2_{\epb} \Ib)$ in a
probit model. The variance of the error term is indeed
\begin{equation}\label{eq:variance}
var(\eb)=var\left[(\Ib-\rho\Wb)^{-1}\epb\right]=\sigma^2_\epb\left[(\Ib-\rho\Wb)'(\Ib-\rho\Wb)\right]^{-1}.
\end{equation}
Let
\begin{equation}\label{eq:sd}
\Db=diag(\ssb_{\eb})
\end{equation}
be the diagonal matrix with diagonal elements $\ssb_{\eb}$
that represent the root square of the diagonal elements in
the matrix (\ref{eq:variance}) and
\begin{equation}\label{eq:stanadization}
\qb=\Db^{-1}(\Ib-\rho\Wb)^{-1}\Xb\bb.
\end{equation}

Since $\bb$ and $\sigma^2_{\epsilon}$ cannot both be
estimated in probit models, \citet{McMillen1992} assumes
$\sigma^2_{\epsilon}=1$.  In the E-step, the observed
dependent variable is replaced by the expectation of the
latent variable $\Yb^*$ conditional on the observed dependent
variable $\Yb$, making use of generalized residuals
\citep{CoxSnell1968, ChesherIrish1987}. To compute this
expectation in the first iteration, the starting values of
the parameters $\bb$ and $\rho$ are used, in subsequent
iterations, the estimated parameters. By computing the
conditional expectation of equation (\ref{eq:SAL}), in
the E-step the following result is used
\begin{equation}
E[\Yb^*/\Yb=\yb]=(\Ib-\rho\Wb)^{-1}\Xb\bb+\Db\frac{\phi_n(\qb)\left[\yb-\Phi_n(\qb)\right]}
{\Phi_n(\qb)\left[1-\Phi_n(\qb)\right]},\label{eq:conditional expectation}\\
\end{equation}
where $\phi_n(\cdot)$ and $\Phi_n(\cdot)$ denote respectively the
$n$-dimensional multivariate probability density and cumulative
distribution functions of a standard normal.

Subsequently, setting $\sigma^2=1$ in the M-step, new estimates are
obtained by maximizing the log-likelihood function
\[k-\frac{1}{2}[(\Ib-\rho\Wb)\yb^*-\Xb\bb]'[(\Ib-\rho\Wb)\yb^*-\Xb\bb]+\sum_{i=1}^nln(1-\rho
\omega_i),\] where $\omega_i$ are the eigenvalues of $\Wb$.
$\prod_{i=1}^n(1-\rho\omega_i)$ is a computationally efficient
approximation of the determinant $|\Ib-\rho\Wb|$ \citep{Ord1975}.
This process is repeated until convergence.\footnote{To obtain a
$\rho$ estimate in the interval (-1,1), we apply the one-to-one
transformation $\rho=-1+2\Phi_1(\rho^*)$, making use of the
invariance of maximum likelihood estimators \citep[p.
253--255]{DavidsonMacKinnon1993}.}

The main advantage of this methodology is that it avoids to compute
an $n$-dimensional integral. The cost is that the E-step requires
the calculation of the inverse of the matrix $(\Ib-\rho\Wb)$.
Although this can be made slightly more efficient by using the
eigenvalues of $\Wb$ to approximate the inverse, it still slows down
the algorithm considerably.  In addition to the computational burden
in the implementation of the algorithm, the main drawback of this
proposal is the covariance matrix estimate of dimensions $n \times
n$. By considering the spatial probit model as a non-linear weighted
least squares model, \citet{McMillen1992} obtains biased but
consistent estimates of the covariance matrix.  For this reason
\citet{McMillen1995b} explores computationally simpler alternatives
to the methods in \citet{McMillen1992}, expressing the belief that
the methods proposed in \citet{McMillen1992} are impractical for
large sample sizes. Another problem with \citet{McMillen1992}'s
approach is the need to specify a functional form for the
nonconstant variance over space \citep{LeSage2000}. In larger models
a practitioner would need to devote considerable effort to testing
the functional form and variables involved in the model for the
variance of the noise elements $\epsilon_i$. Finally, the EM
approach cannot provide an estimate of precision for the spatial
autoregressive parameter $\rho$.

\subsection{Gibbs sampling}

The Gibbs sampler is a particular Markov Chain Monte Carlo (MCMC) introduced by
\citet{GemanGeman1984} in the context of image restoration. When a direct
specification of a joint distribution is not feasible, the Gibbs sampling
procedure specifies the complete conditional distributions for all parameters
in the model and proceeds to sample from these distributions to collect a large
set of parameter draws.  During sampling, a conditional distribution for the
latent observations $y_i^*$ conditional on all other parameters in the model is
considered.\footnote{\citet{GelfandSmith1990} demonstrate that Gibbs sampling
from the sequence of complete conditional distributions for all parameters in
the model produces a set of estimates that converges in the limit to the true
posterior distribution of the parameters.} This distribution is used to produce
a random draw for all $y_i^*$ in the probit model. The conditional distribution
for the latent variables takes the form of a normal distribution centered on
    the predicted value truncated at the left at 0 if $y_i=1$ and at the right at
    1 if $y_i=0$.

The Bayesian Gibbs sampler approach to estimating spatial discrete choice
models (both BSAR and BSEM models) is proposed by \citet{LeSage2000} and is an
extension of the Gibbs sampling method suggested by
\citet{GemanGeman1984}.\footnote{\citet{BolducFortinGordon1997} take a similar
approach for the closely related spatial ordinal probit model.}   This method
exhibits a similarity to the EM algorithm, where the latent unobserved
observations on the dependent variable $y_i^*$ are replaced by estimated
values. The Bayesian approach is different in the way it formulates the
likelihood function and the estimates of the unobserved latent variable.  The
Gibbs estimator remedies the two limitations of \citet{McMillen1992}'s EM
estimator, its slow convergence and its bias in the estimation of standard
errors.

It is important to underline that \citet{LeSage2000} relaxes the assumption of
homogeneity for the disturbances $\epb$ used in BSAR and BSEM models. This
means that the error term $\epb$ can follow a multivariate normal distribution
$\epb\sim N_n(\mathbf{0},\sigma^2_{\epb} \Vb)$ in a probit model, where
$\Vb=diag(v_1,v_2,\dots,v_n)$ and $v_i$ with $i=1,2,\dots,n$ are the variance
parameters to be estimated.  \citet{Greene2003} points out that accounting for
heteroskedasticity is important for probit models because the estimates are
inconsistent in the presence of nonconstant disturbance variances.

In order to assign the priors of a BSAR model,
\citet{LeSage2000} assumes that the priors are independent
\[\pi(\rho,\bb,\sigma,\Vb)=\pi(\rho)\pi(\bb)\pi(\sigma)\pi(\Vb),\]
where
\begin{eqnarray*}
\pi(\rho)&\varpropto & constant\\
\pi\left(v^{-1}_i/q\right)&\backsim & \frac{\chi^2(q)}{q} \qquad i=1,2,..,n\\
\pi(\sigma^2)&\varpropto &\frac{1}{\sigma}.
\end{eqnarray*}

The parameter $q$ controls the amount of dispersion in $v_i$, with small values
of $q$ producing leptokurtic distributions and large values imposing
homoskedasticity.\footnote{$q=7$ produces estimates similar to logit and use of
a large value, e.g. q=100, produces estimates similar to those from probit.} We
summarize \citet{LeSage2000}'s algorithm by the following steps:\footnote{We
follow \citet{Thomas2007}'s implementation of \citet{LeSage2000}'s methodology,
which follows the suggestion by  \citet[p. 159]{Fleming2004} to transform the
latent variable into one that is distibuted independently by using the Cholesky
root of the inverted error covariance matrices (cf. matrix ${\mathbf A}$ in the
RIS estimator below).}

\begin{enumerate}

    \item Initial values for the parameters $\rho_0$, $\bb_0$,
    $\sigma_0$, $\Vb_0$ are considered. The residuals $\epb_0$
    are computed by substituting these values in equation
    (\ref{eq:SAL}). Using a random draw from $\chi^2(n)$ the
    following value is determined:
    \[\sigma^2_1=\frac{\epb_0'\Vb_0\epb_0}{\chi^2(n)}.\]

    \item Given the values $\rho_0$, $\Vb_0$, $\sigma_1$, the
    parameter $\bb_1$ is drawn from the multivariate normal
    \[f(\bb_1/ \rho_0, \Vb_0, \sigma_1)\sim N_n
    \left[(\Xb'\Vb_0^{-1}\Xb)^{-1}\Xb'\Vb_0^{-1}(\Ib-\rho\Wb)\yb^*,\sigma_1^2(\Xb'\Vb_0^{-1}\Xb)^{-1}\right].\]

    \item By drawing an $n$-vector of random $\chi^2(q+1)$ and
    by using $\rho_0$, $\bb_1$, and $\sigma_1$, the values
    $v_i$ with $i=1,2,\dots,n$ are computed with
    \[v_i=\frac{\sigma_1^{-2}\epsilon_{i1}^2+q}{\chi^2(q+1)}.\]

    \item By knowing the values $\bb_1$, $\sigma_1$, $\Vb_1$,
    the \emph{metropolis sampling} algorithm
    \citep{Hastings1970} is applied to determine $\rho_1$. The
    conditional posterior for $\rho$ given $\bb_1$, $\sigma_1$,
    $\Vb_1$ is \begin{equation} \label{eq:metropolis}
    f(\rho_0/\bb_1, \sigma_1, \Vb_1)\propto
    |\Ib-\rho\Wb|exp\left\{-\frac{1}{2\sigma^2}\epb'_1\Vb_1^{-1}\epb_1\right\}.
    \end{equation} Let the value $\rho^*=\rho_0+cZ$ be
    generated, where $Z$ is  a draw from a standard normal
    distribution and $c$ is a known constant.\footnote{For the
    Monte Carlo simulations in this article we set $c=0.1$.}
    The acceptance probability
    $p=min\{1,\frac{f(\rho^*)}{f(\rho_0)}\}$, where $f(\cdot)$
    is defined in equation (\ref{eq:metropolis}). A value
    $m$ is drawn from a continuous uniform distribution with
    support $[0,1]$. If $m<p$, the next draw from the density
    function (\ref{eq:metropolis}) is given by $\rho_1=\rho^*$,
    otherwise the draw is taken to be the current value
    $\rho_1=\rho_0$.

    \item The values of the latent dependent variable $\yb^*$
    are sampled from the multivariate truncated normal
    distribution \[ \Yb^*\sim
    N_n^T((\Ib-\rho_1\Wb)^{-1}\Xb\bb_1, \bm{\Lambda}),\] where
    $\bm{\Lambda}$ is the diagonal matrix whose elements are
    the elements of the main diagonal of the matrix
    $(\Ib-\rho_1\Wb)^{-1}\epb\epb'[(\Ib-\rho_1\Wb)^{-1}]'.$ The
    normal distribution is truncated at the left at 0 if $Y=1$
    and truncated at the right at 0 if $Y=0$
    \citep{AlbertChib1993}.

\end{enumerate}

\citet{LeSage2000}'s approach overcomes the problems in estimating the standard
error in the EM algorithm since parameter standard errors are derived from the
posterior parameter distributions. The first advantage of the Bayesian strategy
is to be able to derive the condition distribution of each parameter, and thus
compute different moments of the distribution. The second advantage is its
flexibility to account for the heteroskedasticity in the error terms.

\subsection{Recursive Importance Sampling}

\citet{BeronVijverberg2004} propose a recursive importance sampling (RIS)
estimator to evaluate directly the $n$-dimensional integral in both the BSAR
and the BSEM models. The RIS-normal simulator is identical to what is sometimes
called the Geweke-Hajivassiliou-Keane (GHK) simulator
\citep{BorschSupanHajivassiliou1993}.

Define $\mathbf{Z}$ as an $n\times n$ matrix with
\[z_{ij}=\left\{
           \begin{array}{ll}
             1-2y_i & \text{if}\quad i=j \\
             0 & \text{otherwise,}
           \end{array}
         \right.
\]
for $i,j=1,2,\dots,n$. This means that $\mathbf{Z}$ is a diagonal
matrix that satisfies the equation $\mathbf{Z}\mathbf{Z}'=\Ib_n$. By
defining $\mathbf{t}=\mathbf{Z}\mathbf{e}$,  we obtain that
\[Var(\mathbf{t})=\mathbf{Z}Var(\mathbf{e})\mathbf{Z}'\equiv \bm{\Sigma}_{\rho}, \]
where $Var(\mathbf{e})$ is provided by equation
(\ref{eq:variance}). By this notation the observed vector
$\mathbf{y}$ defined in (\ref{eq:observed variable}) leads to an
upper limit on $\mathbf{t}$:
\[\mathbf{t}<-\mathbf{Z}(\Ib-\rho\Wb)^{-1}\Xb\bb,\]
which means that we can write the log-likelihood function as
\begin{equation}\label{eq:likelihood}
\ln L=\ln\Phi_n\left[-\mathbf{Z}(\Ib-\rho\Wb)^{-1}\Xb\bb;\mathbf{0},\bm{\Sigma}_{\rho}\right]\equiv\ln\Phi_n\left[\bm{T};\mathbf{0},\bm{\Sigma}_{\rho}\right],
\end{equation}
where $\Phi_n\left[\bm{j};\bm{\mu},\bm{\Omega}\right]$ is a
$n$-dimensional normal cumulative distribution function with mean
vector $\mathbf{\mu}$ and
variance-covariance matrix $\bm{\Omega}$.

In order to evaluate the probability in equation (\ref{eq:likelihood}),
\citet{BeronVijverberg2004} propose to apply the RIS simulator, developed in
detail by \citet{Vijverberg1997}.  Let $\mathbf{A}$ be an upper triangular
matrix such that $\mathbf{A}'\mathbf{A}=\Sigma_{\rho}^{-1}$ and let
$\mathbf{\eta}=\mathbf{A}\mathbf{t}$. Whether $\Sigma_{\rho}$ is standardized
or not, the vector $\mathbf{\eta}$ is i.i.d. standard normal. By defining the
matrix $\mathbf{B}=\mathbf{A}^{-1}$, $\mathbf{B}$ results an upper triangular
matrix with $b_{jj}>0$ $\forall j$ and $\mathbf{B}\mathbf{\eta}=\mathbf{t}$.

Given the upper bound
$\{\mathbf{B}\mathbf{\eta}=\mathbf{t}\}<\mathbf{T}$, we can apply
the following iterative procedure:
\begin{eqnarray}
\eta_n&<&b^{-1}_{nn}T_n\equiv \eta_{n0}\nonumber\\
\eta_j&<&b^{-1}_{jj}\left[T_j-\sum_{i=j+1}^n{b_{ji}}\eta_i\right]\equiv\eta_{j0}(T_j,\eta_{j+1},\ldots,\eta_n)\equiv\eta_{j0}.\label{eq:upper
bounds}
\end{eqnarray}
Let $g(\eta_j)$ be a probability density function with support the
whole real axis and let $G(\cdot)$ be the associated cumulative
distribution function. By denoting
\begin{equation*}
g^c(\eta_j)=\frac{g(\eta_j)}{G(\eta_{j0})}
\end{equation*}
for $\eta_j\leq\eta_{j0}$, we can compute the following probability:
\begin{equation}\label{probability}
\begin{split}
p&=P\{\mathbf{t}<\mathbf{T}\}=\int_{-\infty}^\mathbf{T}
\phi(\mathbf{t};\mathbf{0},\bm{\Omega})d\mathbf{t}=\int_{-\infty}^{\eta_{n0}}\ldots\int_{-\infty}^{\eta_{1,0}}\prod_{j=1}^n
\phi(\eta_j)d\eta_1\ldots d\eta_n\\
&=\int_{-\infty}^{\eta_{n0}}\frac{\phi(\eta_n)}{g^c(\eta_n)}\left[\int_{-\infty}^{\eta_{n-1,0}}\frac{\phi(\eta_{n-1})}{g^c(\eta_{n-1})}\ldots
\left(\int_{-\infty}^{\eta_{2,0}}\frac{\phi(\eta_2)}{g^c(\eta_2)}\Phi(\eta_{10})g^c(\eta_2)d\eta_2\right)\ldots\right]g^c(\eta_n)d\eta_n.
\end{split}
\end{equation}

The RIS simulator is implemented by drawing a large number $R$ of
random vector $\bm{\eta}$ satisfying the condition
$\eta_j\leq\eta_{j0}$ for $j=1,2,\dots,n$ from the density function
$g(\cdot)$.\footnote{For the
    Monte Carlo simulations we use $R=1000$.}  There are different suitable
    density functions used to define $g(\cdot)$ \citep{Vijverberg1997}.
    \citet{Vijverberg1999} shows that the RIS-normal simulator is often
    preferred. For this reason we choose the normal density function in the
    following Monte Carlo simulations and, in particular, we apply the
    antithetical sampling strategy suggested by \citet{Vijverberg1997} for
    simulating from a multivariate normal distribution.

The recursive nature of the RIS simulator is due to the fact that the bounds in
equation (\ref{eq:upper bounds}) are backwards determined. For every drawing
$r$ of the random vector $\bm{\eta}$, given $\eta_{n0}$, the values
$\tilde{\eta}_{n,r}$ and $\tilde{\eta}_{n-1,0,r}$ are calculated using equation
(\ref{eq:upper bounds}) by using $\tilde{\eta}_{n,r}$ in the place of $\eta_n$.
This process is repeating until $\tilde{\eta}_{1,0,r}$ is computed. Then for
the RIS-normal simulator the simulated value for $p$, defined in equation
(\ref{probability}), is
\[\hat{p}=\frac{1}{R}\sum_{r=1}^R\left(\prod_{j=1}^n\Phi[\tilde{\eta}_{j,0,r}]\right),\]
where $\Phi(\cdot)$ is the cumulative distribution function of the one
dimensional standard normal random variable.

Based on the Monte Carlo study that \citet{BeronVijverberg2004} performed the
RIS simulator can provide accurate estimates for spatial binary choice models.
Moreover, this approach is attractive since it is the only one that directly
evaluates the $n$-dimensional probit likelihood function. This means that only
this methodology allows for the use of the Likelihood Ratio test. Because of
these advantages \citet{BeronMurdochVijverberg2003} and \citet{Novo2003} apply
the RIS simulator.

\subsection{Generalized Method of Moments}

This section describes a spatially dependent binary choice methodology that
considers the problem as a weighted non-linear version of the linear
probability \citep{Amemiya1985,Greene2002,Maddala1983} with a
variance-covariance matrix that can be estimated with a Generalized Method of
Moments (GMM) estimator \citep{Hansen1982}.  \citet{PinkseSlade1998} derive the
GMM moment equations from the likelihood function. \citet{KlierMcMillen2008}
propose a linearized version of the GMM suggested by \citet{PinkseSlade1998}.

\subsubsection{Pinkse and Slade's estimator}

While \citet{PinkseSlade1998} suggest to apply the GMM to a BSEM model, for
achieving the aim of this article we present their estimator for a BSAR model.
Similar to \citet{McMillen1992}, \citet{PinkseSlade1998} consider the
generalized residuals\footnote{Unlike \citet{ChesherIrish1987} and
\citet{CoxSnell1968} (see also eq. (\ref{eq:conditional expectation})),
\citet{PinkseSlade1998} define the generalized residuals as $\Db^{-1}E[\eb/
\yb,\eb,\rho]$ and not $E[\eb/\yb,\eb,\rho]$.}

\begin{equation}\label{eq:generalized residuals2}
\tilde{\eb}(\bm{\theta})=\Db^{-1}E[\eb/\yb,\bm{\theta}]=
\frac{\phi_n\left[\qb(\bm{\theta})\right]\left\{\yb-\Phi_n[\qb(\bm{\theta})]\right\}}{\Phi_n\left[\qb(\bm{\theta})\right]\left\{1-\Phi_n[\qb(\bm{\theta})]\right\}},
\end{equation}
where $\bm{\theta}=(\bb',\rho)'$ is the parameter vector and $\Db$ and
$\qb$ are defined in equations (\ref{eq:sd}) and
(\ref{eq:stanadization}), respectively.

By applying the GMM the parameter vector $\bm{\theta}$ is estimated
by
\begin{equation}\label{eq:moments}
\hat{\bm{\theta}}=\arg\min_{\bm{\theta}\in\bm{\Theta}}\:
\tilde{\eb}'(\bm{\theta})\Zb\Mb \Zb'\tilde{\eb}(\bm{\theta}),
\end{equation}
 where $\tilde{\eb}$ is defined in equation (\ref{eq:generalized residuals2}), $\Zb$ is
 a matrix of instruments,\footnote{In the Monte Carlo simulations we consider
 $\Zb=\mathbf{1}+\Xb+\Wb\Xb+\Wb^2\Xb+\Wb^3\Xb$.} $\Mb$ is a positive definite
 matrix\footnote{\citet{PinkseSlade1998} consider $\Mb$ equal to the identity
 matrix $\Mb=\mathbf{I}_n$ in their empirical application and we follow this
 suggestion in the Monte Carlo simulations.} and $\bm{\Theta}$ is the
 parametric space.

\citet{PinkseSlade1998} provide the asymptotic variance of their
estimator for a BSEM model and develop also the hypothesis test for
spatial error correlation. Their approach overcomes the problems of
evaluating a high order integral and the $n$ by $n$ determinants in
the Maximum Likelihood method. The main disadvantage of this approach is that it
requires the $n\times n$ matrix $(\Ib-\rho\Wb)^{-1}$ to be inverted
in each iteration. Furthermore, since \citet{PinkseSlade1998} apply
the GMM method, their estimator is less efficient than the ML
estimators.

\subsubsection{Klier and McMillen's estimator}

\citet{KlierMcMillen2008} linearize \citet{PinkseSlade1998}'s model
around a convenient starting point for a BSAR logit model. In
particular, in equation (\ref{eq:moments}),
\citet{KlierMcMillen2008} let $\Mb=(\Zb'\Zb)^{-1}$, so the objective
function for the GMM estimator is
\begin{equation*}
\hat{\bm{\theta}}=\arg\min_{\bm{\theta}\in\bm{\Theta}}\:
\tilde{\eb}'(\bm{\theta})\Zb(\Zb'\Zb)^{-1}
\Zb'\tilde{\eb}(\bm{\theta}),
\end{equation*}
hence \citet{KlierMcMillen2008} apply a nonlinear two-stage least
squares method. In order to analyse  \citet{KlierMcMillen2008}'s methodology we
define
\begin{equation}
\mathbf{P}=P\{\Yb=1/\bm{\theta}\}=\frac{\exp[\qb(\bm{\theta})]}{1+\exp[\qb(\bm{\theta})]}.
\end{equation}
where $\qb(\bm{\theta})$ is defined in equation
(\ref{eq:stanadization}).

 \citet{KlierMcMillen2008}'s iterative procedure has the
following steps:
\begin{enumerate}
  \item assume initial values for the parameter vector
$\bm{\theta}_0=(\bb_0',\rho_0)'$;
  \item compute $\eb_0$ defined in equation (\ref{eq:generalized residuals});
\item compute the gradient terms
\begin{eqnarray}
\Gb_{\bb i }&=&\frac{\partial P_i}{\partial \bb} =
\hat{P}_i(1-\hat{P}_i)\mathbf{t}_i \nonumber\\
G_{\rho i}&=&\frac{\partial
P_i}{\partial\rho}=\hat{P}_i(1-\hat{P}_i)\left[h_i-\frac{q_i}{\sigma^2_{ei}}\upb_{ii}\right], \label{eq:km_deriv}
\end{eqnarray}
where $\mathbf{t}_i$ is the $i$-th row vector of the matrix
$\mathbf{T}=\Db^{-1}(\Ib-\rho\Wb)^{-1}\Xb$, $h_i$ is the $i$-th element of the
vector $\mathbf{h}=(\Ib-\rho \Wb)^{-1} \Wb \qb$, $q_i$ is the $i$-th element of
the vector $\qb$ defined in equation (\ref{eq:stanadization}) and $\Upsilon_{ii}$
is the $i$-th element of the diagonal of the matrix $\upb=(\Ib-\rho \Wb)^{-1}
\Wb (\Ib-\rho \Wb)^{-1} (\Ib-\rho \Wb)^{-1}$.\footnote{The derivative in equation
(\ref{eq:km_deriv}) assumes that $(\Ib-\rho\Wb)$ is a symmetric matrix, which
is not guaranteed. For example, when $\Wb$ is standardized, this is not the case. See the appendix for this
derivative without assuming a symmetric matrix. The revised derivative in the
appendix  does not affect the estimator or the Monte Carlo results when the convenient starting point at $\rho=0$ is used, as in \citet{KlierMcMillen2008}.}
\item regress the gradient terms $\Gb_{\bb }$ and $G_{\rho}$ on
$\Zb$ and compute the predicted values $\hat{\Gb}_{\bb }$ and
$\hat{G}_{\rho }$;
\item regress $\eb_0+\Gb_{\bb }\hat{\bb_0}$ on $\hat{\Gb}_{\bb }$
and $\hat{G}_{\rho }$. The coefficients obtained from this
regression are the estimated values of $\bb$ and $\rho$.
\end{enumerate}

The main advantage of this approach is that it is not iterative and
does not require the $n\times n$ matrix $(\Ib-\rho \Wb)^{-1}$ to be
inverted in each iteration, unlike \citet{PinkseSlade1998}'s
estimator. This characteristic leads to a computationally
significantly faster estimator. The main disadvantage of this
estimator is that it provides accurate estimates of $\rho$ only as
long as $\rho$ is small.  Furthermore, linearized approach cannot
provide an estimate of precision for the spatial autoregressive
parameter $\rho$. Finally, since \citet{KlierMcMillen2008} propose a
linearization around the starting point, a restriction for the
parameter $\rho$ to the interval [-1,1] cannot be introduced by
using their method.

\section{Monte Carlo simulations}

In order to make up for the lack of simulation studies for BSAR models
\citep{Fleming2004}, in this section we compare the properties of these five
estimators by Monte Carlo simulations. The set up of the simulations is
primarily based on the literature on policy and regime diffusion \citep[e.g.][]{GleditschWard2006, BaturoGray2009} and on broad
similarity with simulations as published in accompaniment of the proposals of
estimators discussed in this paper \citep[e.g.][]{McMillen1995, BeronVijverberg2004, KlierMcMillen2008}. In order to understand how the properties
of these estimators vary according to the number of observations, we consider
two different sample sizes: $n=50$ and $n=500$. The first sample size is set
because it resembles the number of states in the US, which is a typical
application area for studies in policy diffusion \citep[e.g.,][]{Mooney2001,
Volden2006, LeSageParent2007}. The larger sample size is added to be able to
see the results when the sample size increases.

In our Monte Carlo analysis we generate 1,000
replications.\footnote{The same number of replications is considered
by \citet{FloresLagunesSchnier2005, FranzeseHays2007,
KlierMcMillen2008}.} For generating the data sets\footnote{We use
the {\tt R} package {\tt rlecuyer} for the parallel generation of
random numbers.} we consider one covariate $X$ drawn from a normal
distribution $N(2,4)$ with expected value 2 and standard deviation
4.\footnote{Following \citet{McMillen1995}, we prefer to consider a
standard deviation of $X$ substantially higher than $\sigma_{\epb}$.} Based on
equation (\ref{eq:SAL}), the residuals vector $\epb$ is generated
from a multivariate normal distribution $N_n(\vec{0}, \Ib)$ and the
parameter vector is $\bb=[4,-2]'$. In order to generate $\Wb$, we
apply the method suggested by \citet[p. 179]{BeronVijverberg2004} by
using $d=0.21$ for $n=50$ and $d=0.06$ for $n=500$. To analyse how
the characteristics of the estimators change according to the level
of autocorrelation, we consider four different values
$(0;0.1;0.45;0.8)$ of the parameter $\rho$, such that the last three
values are equidistant.\footnote{\citet{Anselin1982},
\citet{BeronVijverberg2004} and \citet{KlierMcMillen2008} consider
similar values.} For the maximization procedure in the EM, RIS, and
\citet{PinkseSlade1998} estimators, we use the {\tt optim()}
function in {\tt R} with a maximum number of iterations of 1,000.
Finally, analogously to \citet{LeSage2000} and
\citet{BeronVijverberg2004} we consider a maximum number of loops
equal to 1,000 for the RIS and EM algorithms, and 3,000 for the
Gibbs estimator.

In the following tables we report the mean of the bias and the standard
deviation of the estimators (in round brackets) computed on 1,000 sets. The
data is generated based on a probit model. Since \citet{KlierMcMillen2008} have
proposed their estimator for the logit model, we rescale their parameter
estimates to allow for comparison. Because the variance of the logistic
distribution is $\pi^2/3$, we report the estimates $\hat{\beta}_0\sqrt{3}/\pi$
and $\hat{\beta}_1\sqrt{3}/\pi$ for the linearized GMM estimator.\footnote{The
spatial coefficient $\rho$ is not affected by the scaling.}

\begin{figure}[ht!]
\includegraphics[width=\textwidth]{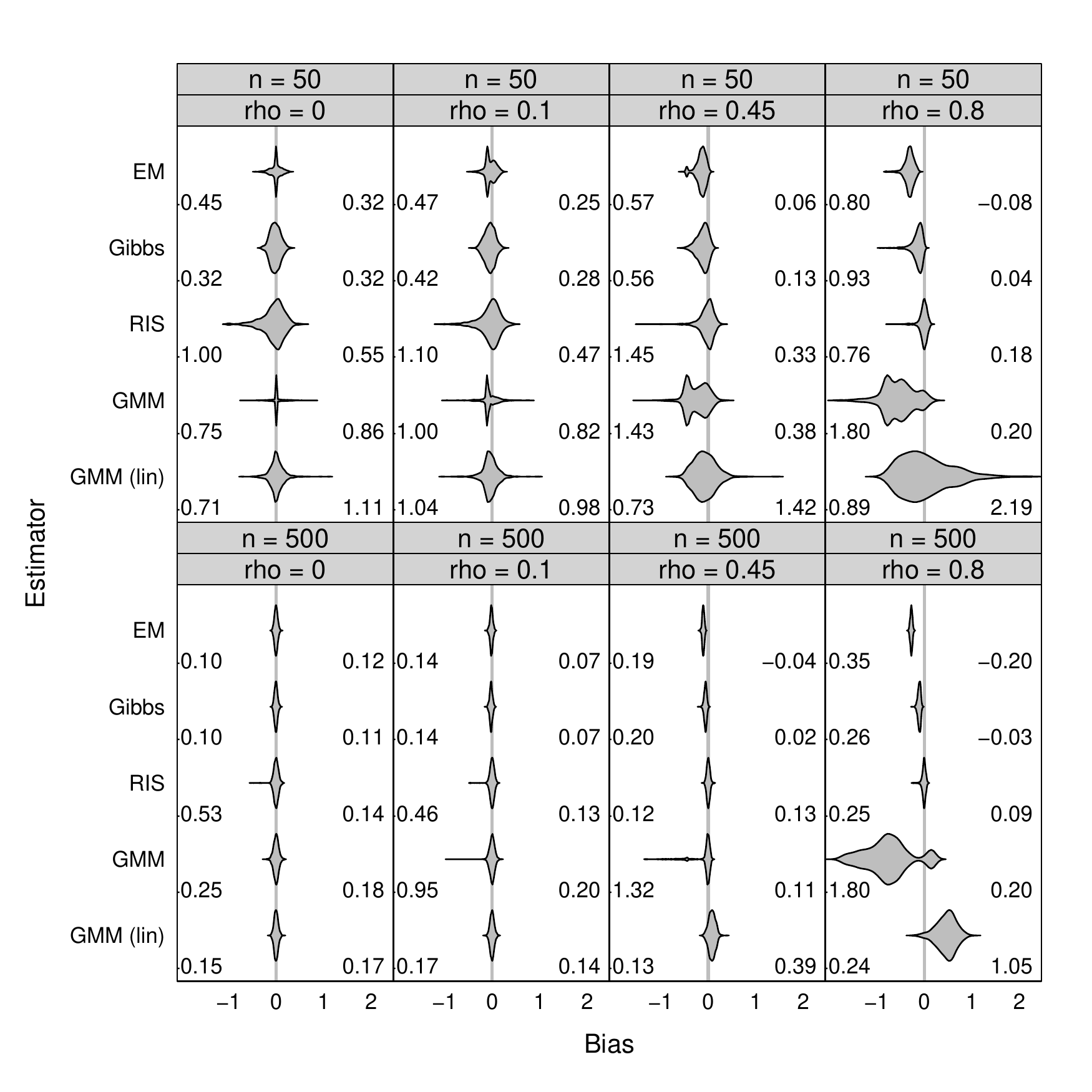}
\caption{\small The distribution of the bias of the estimators of
the autocorrelation parameter $\rho$ obtained from Monte Carlo
simulations on 1,000 samples. Numbers represent minimum and maximum
values of the bias.} \label{fig:rhobias}
\end{figure}

The primary focus of this paper is on the estimation of the level of
spatial autocorrelation in a binary spatial autoregressive model
specification. Since we have the study of the
diffusion of policies and regimes in mind, the level of diffusion is
typically of key interest. The autocorrelation in the residuals is
thus not treated as a mere nuisance, but as a structural factor of
substantive interest. Figure \ref{fig:rhobias} provides the
distribution of the bias ofthe estimators of the spatial autocorrelation parameter
$\rho$ in the BSAR model described in equation
(\ref{eq:SAL}).\footnote{These plots are generated using the {\tt
violin} function in the {\tt lattice} package in {\tt R}, which in
turn makes use of the built-in {\tt density} function for the
computation of smoothed kernel density estimates. Default settings
are used.} Table \ref{tab:rhobias} provides the mean and standard
deviation of the bias of the above-mentioned estimators..

\begin{table}[ht!]
\begin{center}
\begin{tabular}{|l|cc|cc|cc|cc|}
\cline{2-9}
\multicolumn{1}{c|}{} & \multicolumn{2}{c}{\textbf{$\rho=0$}} & \multicolumn{2}{|c|}{\textbf{$\rho=0.1$}} & \multicolumn{2}{|c|}{\textbf{$\rho=0.45$}} & \multicolumn{2}{|c|}{\textbf{$\rho=0.8$}}\\
\hline
\backslashbox{$\hat{\rho}$}{$n$} & \textbf{50} & \textbf{500} & \textbf{50} & \textbf{500}& \textbf{50} & \textbf{500} & \textbf{50} & \textbf{500}  \\
\hline
\multirow{2}{*}{\emph{\textbf{EM}}}
 & -0.002 & -0.002 & -0.038 & -0.016 & -0.148 & -0.104 & -0.311 & -0.272\\
 & (0.108) & (0.034) & (0.108) & (0.030) & (0.107) & (0.022) & (0.098) & (0.023)\\
\hline
\multirow{2}{*}{\emph{\textbf{Gibbs}}}
 & -0.017 & -0.003 & -0.050 & -0.023 & -0.114 & -0.058 & -0.140 & -0.104\\
 & (0.108) & (0.032) & (0.109) & (0.030) & (0.120) & (0.028) & (0.117) & (0.030)\\
\hline
\multirow{2}{*}{\emph{\textbf{RIS}}}
 & -0.051 & -0.002 & -0.044 & -0.003 & -0.027 & 0.003 & -0.015 & -0.005\\
 & (0.235) & (0.056) & (0.219) & (0.053) & (0.158) & (0.034) & (0.093) & (0.039)\\
\hline
\multirow{2}{*}{\emph{\textbf{GMM}}}
 & 0.019 & -0.001 & -0.049 & -0.001 & -0.243 & -0.097 & -0.547 & -0.775\\
 & (0.165) & (0.055) & (0.173) & (0.060) & (0.228) & (0.245) & (0.336) & (0.458)\\
\hline
\multirow{2}{*}{\emph{\textbf{GMM {\footnotesize (lin)}}}}
 & 0.011 & -0.001 & -0.037 & -0.001 & -0.075 & 0.080 & 0.028 & 0.468\\
 & (0.149) & (0.043) & (0.166) & (0.044) & (0.232) & (0.072) & (0.528) & (0.205)\\
\hline
\end{tabular}

\end{center}
\caption{\small The mean bias and the standard deviation (between
parentheses) of the estimators of the autocorrelation parameter
$\rho$ obtained from Monte Carlo simulations on 1,000 samples.}
\label{tab:rhobias}
\end{table}

It is clear from Figure \ref{fig:rhobias} that the performance of
the estimators varies depending on the level of autocorrelation in the
data, with particularly large differences between estimators under
high levels of autocorrelation. As can be seen in Table
\ref{tab:rhobias}, in the absence of spatial autocorrelation
($\rho=0$), the Gibbs and the EM estimators are the best estimators
of $\rho$ in terms of both the distortion and the dispersion. The
linearized GMM estimator also does particularly well, which is
unsurprising, since $\rho=0$ is the value used as the starting point
for the linearization. When looking at the distribution as a whole,
however, it is clear that while this estimator performs generally
well, there are clear outliers among the estimates. The RIS
estimator shows the worst performance and it tends to underestimate
$\rho$ for both small ($n=50$) and large ($n=500$) samples, with in
particular some negative outliers. The \citet{PinkseSlade1998}
estimator, on the other hand, tends to overestimate $\rho$ in this
scenario.

When the level of spatial autocorrelation is positive but still limited
($\rho=0.1$), the EM and Gibbs estimators still show good performance, even if
the differences with the other estimators are less considerable in comparison
with the scenario without spatial autocorrelation. Both the EM and the Gibbs
estimators tend to underestimate the spatial autocorrelation parameter. As can
be expected, the linearized GMM estimator is still performing well this close
to the linearization point of $\rho=0$. The main disadvantage of this estimator
is that it is not possible to put a reasonable constraint on $\hat{\rho}$, such
that occasional estimates are obtained outside the $[-1,1]$ interval, visible
in Figure \ref{fig:rhobias}. The RIS estimator is also prone to the occasional
outlier in its estimate of the level of autocorrelation. It is striking that
all estimators, with the exception of GMM with $n=50$, tend to underestimate
$\rho$, often by over 50\% of the true parameter value, when the sample size is
low.

The next scenario contains significant levels of autocorrelation,
with $\rho=0.45$. Under this level of autocorrelation, the RIS
estimator starts to perform relatively well compared to the other
estimators. While the absolute mean bias is the lowest, the
variation is still relatively high, however, and the plot clearly
shows the presence of some outliers. For the EM and the Gibbs
estimators, the bias is significantly larger than for $\rho=0.1$ and
for the linearized GMM, a clear increase in the dispersion is visible
in Figure \ref{fig:rhobias}, although this is compensated by a
reduction in extreme outliers. The distribution of the bias of the
GMM estimator now shows a clear tendency to underestimate the amount
of autocorrelation, with a tight distribution under $n=500$, but
with some outliers.

For high spatial autocorrelation ($\rho=0.8$), the RIS estimator
shows the best performance. The
linearized GMM now clearly suffers from the
large distance from the starting point of the linearization -- the
extrapolation from $\rho=0$ to $\rho=0.8$ leads to significant
overestimation of the level of autocorrelation. The plot also shows
that the estimator clearly suffers from the lack of a constraint on
$\rho$. The GMM estimator shows rather significant underestimation
of $\rho$, as well as a high level of variance. Furthermore, under
this scenario, the simulations suggest asymptotically biased in mean
results for the GMM and the linearized GMM estimators, with
the mean bias increasing for the larger sample
size.\footnote{
When the observations are ``strongly spatially dependent'' \citep[p. 134, fn.
12]{PinkseSlade1998}, even the consistency  is  not guaranteed.}
Following the trend already visible when moderately increasing $\rho$,
the EM estimator clearly shows greater mean bias under this
scenario, while for Gibbs, the results are similar to $\rho=0.45$.

\begin{sidewaystable}
\begin{center}
\begin{tabular}{|l|cc|cc|cc|cc|}
\cline{2-9}
\multicolumn{1}{c|}{} & \multicolumn{2}{c}{\textbf{$\rho=0$}} & \multicolumn{2}{|c|}{\textbf{$\rho=0.1$}} & \multicolumn{2}{|c|}{\textbf{$\rho=0.45$}} & \multicolumn{2}{|c|}{\textbf{$\rho=0.8$}}\\
\hline
\backslashbox{$\hat{\beta_1}$}{$n$} & \textbf{50} & \textbf{500} & \textbf{50} & \textbf{500}& \textbf{50} & \textbf{500} & \textbf{50} & \textbf{500}  \\
\hline
\multirow{2}{*}{\emph{\textbf{EM}}}
 & -59.23 & -0.12 & -43.12 & -0.02 & -5.52 & 1.03 & 1.42 & 1.69\\
 & (507.24) & (0.34) & (356.69) & (0.33) & (56.98) & (0.11) & (3.64) & (0.03)\\
\hline
\multirow{2}{*}{\emph{\textbf{Gibbs}}}
 & -0.95 & -0.07 & -0.86 & -0.05 & -0.63 & 0.09 & 0.30 & 0.79\\
 & (1.12) & (0.28) & (1.09) & (0.29) & (1.02) & (0.24) & (0.70) & (0.19)\\
\hline
\multirow{2}{*}{\emph{\textbf{RIS}}}
 & -8.05 & 0.01 & -5.86 & 0.02 & -6.86 & 0.13 & -4.13 & 1.12\\
 & (19.89) & (0.38) & (13.06) & (0.40) & (18.59) & (0.43) & (12.62) & (0.49)\\
\hline
\multirow{2}{*}{\emph{\textbf{GMM}}}
 & -5035.95 & -695.99 & -4194.69 & -576.70 & -4777.75 & -17864.07 & -2818.31 & -80663.45\\
 & (28279.94) & (6344.00) & (23060.63) & (5841.52) & (15011.28) & (250693.49) & (11646.26) & (336721.33)\\
\hline
\multirow{2}{*}{\emph{\textbf{GMM {\footnotesize (lin)}}}}
 & -64.15 & -0.04 & -41.08 & 0.07 & -8.32 & 1.13 & 1.25 & 1.70\\
 & (424.80) & (1.65) & (175.62) & (1.93) & (95.09) & (0.49) & (6.82) & (0.05)\\
\hline
\end{tabular}

\end{center}
\caption{\small The mean bias and the standard deviation (between
parentheses) of the estimators of the parameter $\beta_1$ obtained
from Monte Carlo simulations on 1,000 samples.} \label{tab:b1bias}
\end{sidewaystable}

Even with primary focus on the level of autocorrelation, the
estimates of the effects of other independent variables can of
course not be ignored. Table \ref{tab:b1bias} provides the mean and
the standard deviation of the estimates of $\beta_1$, the parameter
for independent variable $X$. The differences in estimate accuracy
between the estimators vary more dramatically than for $\rho$,
with the Gibbs estimator clearly outperforming all other estimators
under all conditions and the estimator proposed by
\citet{PinkseSlade1998} clearly providing the worst results for both
the bias and the dispersion.

Under absence of spatial autocorrelation ($\rho=0$) and for $n=50$, the Gibbs estimator
generates the least average bias on $\hat{\beta_1}$, with RIS,
linearized GMM, and EM also performing well as long as the sample
size is sufficiently large. Under $n=50$, the Gibbs estimator
underestimates $\beta_1$ by about 50\% on average of the value of
$\beta_1$ and the other
estimators well beyond that. The difference between the smaller and
the larger sample sizes is more pronounced than for the estimates of
$\rho$. The GMM estimator is the only estimator that is still
significantly biased when the sample size is reasonably large. All
estimators tend to underestimate $\beta_1$.


Under limited autocorrelation ($\rho=0.1$), the order of accuracy of
the estimators remains more or less the same, with still only Gibbs
performing the best under the small sample size and performing the well
under the large sample size. The RIS performs similar to EM, Gibbs and GMMlin for large sample size.
The GMM estimator shows the worst performance.
Increasing the autocorrelation to $\rho=0.45$, the results for the
Gibbs, the RIS and the GMM estimators are very similar to
$\rho=0.1$, but the EM and the linearized GMM estimators show
lower mean biases under small sample size.

For the estimation of $\rho$, we saw significant difference between
the moderate and the high levels of autocorrelation -- to what
extent is this the case for the slope coefficients? Similar to the
estimates of $\rho$, the GMM estimator of \citet{PinkseSlade1998}
starts to show very significant distortion and dispersion when the
autocorrelation is high and, furthermore, shows a significant
increase in the mean bias when the sample size increases. The
linearized version of this estimator also reflects this increasing
mean bias with sample size for high autocorrelation, but nevertheless performs remarkably
well, with an average bias of approximately 50 to 75\%  of
$\beta_1$, underestimating the magnitude of the negative effect of
$X$. The Gibbs estimator outperforms all other estimators both with
small and large sample sizes, followed by the RIS, the EM and the
linearized GMM estimators as long as sample size is large. It should
be noted that for all estimators, the bias is relatively large,
which is an overestimate of the magnitude by slightly over 10\%.

\begin{sidewaystable}
\begin{center}
\begin{tabular}{|l|cc|cc|cc|cc|}
\cline{2-9}
\multicolumn{1}{c|}{} & \multicolumn{2}{c}{\textbf{$\rho=0$}} & \multicolumn{2}{|c|}{\textbf{$\rho=0.1$}} & \multicolumn{2}{|c|}{\textbf{$\rho=0.45$}} & \multicolumn{2}{|c|}{\textbf{$\rho=0.8$}}\\
\hline
\backslashbox{$\hat{\beta_0}$}{$n$} & \textbf{50} & \textbf{500} & \textbf{50} & \textbf{500}& \textbf{50} & \textbf{500} & \textbf{50} & \textbf{500}  \\
\hline
\multirow{2}{*}{\emph{\textbf{EM}}}
 & 121.85 & 0.24 & 85.58 & 0.05 & 12.52 & -2.06 & -3.02 & -3.39\\
 & (1120.25) & (0.70) & (689.86) & (0.69) & (127.40) & (0.23) & (11.82) & (0.11)\\
\hline
\multirow{2}{*}{\emph{\textbf{Gibbs}}}
 & 1.92 & 0.13 & 1.72 & 0.09 & 1.25 & -0.19 & -0.62 & -1.58\\
 & (2.44) & (0.59) & (2.34) & (0.60) & (2.21) & (0.50) & (1.68) & (0.39)\\
\hline
\multirow{2}{*}{\emph{\textbf{RIS}}}
 & 15.59 & -0.01 & 11.62 & -0.04 & 13.41 & -0.26 & 8.14 & -2.29\\
 & (38.42) & (0.80) & (26.16) & (0.82) & (36.16) & (0.87) & (25.25) & (1.03)\\
\hline
\multirow{2}{*}{\emph{\textbf{GMM}}}
 & 10553.11 & 1497.72 & 8927.67 & 1215.51 & 9761.46 & 21968.93 & 5390.80 & 48233.04\\
 & (57014.77) & (13763.26) & (46734.98) & (12202.29) & (30576.23) & (293214.42) & (26072.56) & (214363.69)\\
\hline
\multirow{2}{*}{\emph{\textbf{GMM {\footnotesize (lin)}}}}
 & 122.91 & 0.09 & 81.80 & -0.13 & 17.71 & -2.26 & -2.72 & -3.41\\
 & (725.66) & (3.30) & (340.99) & (3.88) & (202.43) & (1.01) & (22.23) & (0.22)\\
\hline
\end{tabular}

\end{center}
\caption{\small The mean bias and the standard deviation (between parentheses) of the estimators of the parameter $\beta_0$ obtained from Monte Carlo
simulations on 1,000 samples.} \label{tab:b0bias}
\end{sidewaystable}

Usually of least concern to applied researchers, but relevant for accurate
prediction, is the estimate of the intercept of the model, in this case
$\beta_0$. Table \ref{tab:b0bias} provides the simulation results for this
parameter of the model. Not unsurprisingly, the results are closely
in line with those for $\beta_1$. Indeed, the mean of the bias of $\beta_0$ is
roughly the bias in Table \ref{tab:b1bias} multiplied by a factor $-2$.
Relative to the size of $\beta$, the bias is thus the same on average.
Similarly, the standard deviation of the bias is multiplied by a factor $2$,
with the exception of the GMM estimator under higher levels of autocorrelation,
where the dispersion under larger sample size is similar to that for $\beta_1$.
The same relative results for the different estimators are therefore obtained.

\section{Conclusion}

In this paper we provide a comprehensive overview of estimators for spatial
autoregressive models with binary dependent variables. These models are of
particular interest to various applications in economics, political science,
and related disciplines, where the outcome might be a policy, a decision, a
transition, or otherwise binary outcome. Applications can also be imagined in
the field of bioinformatics or neuroscience, although sample sizes tend to be
magnitudes larger than those studied here. Many of these outcomes are
interdependent through either spatial contiguity or any other form of
proximity, including social networks or economic linkages, and ignoring the
inherent spatial structure of the data generates inconsistent and inefficient
estimates \citep{McMillen1992}. Furthermore, in many applications the
researcher is explicitly concerned with estimating the level of interdependence
-- it is indeed this concern that is of primary interest in our discussion and
simulation study.

This paper compares five estimators introduced to date for this specific type
of model. An extensive simulation study compares the performance of these five
estimators under conditions of relatively small sample sizes and varying levels
of spatial autocorrelation. When taking both the estimation of the extend of
spatial autocorrelation and the coefficients on the other explanatory variables
into account, the Gibbs estimator \citep{LeSage2000} clearly outperforms the
other estimators. When the sample size increases, the difference between the
different estimators becomes smaller. When focusing specifically on the spatial
autoregressive component alone,  the Gibbs estimator \citep{LeSage2000}
performs best for low spatial autocorrelation, while the Recursive Importance
Sampler \citep{BeronVijverberg2004} performs best for high spatial
autocorrelation. The linearized GMM estimator \citep{KlierMcMillen2008} is an
interesting option when the sample size is large and the autocorrelation
relatively low, in particular due to its high computational speed.

\bibliographystyle{apsr}
\bibliography{dissertation}

\appendix

\section*{Appendix: Derivative for the linearization of the GMM BSAR model}

The gradient (\ref{eq:km_deriv}) of interest to the linearization
proposed in \citet{KlierMcMillen2008} is the derivative of the
logistic link function to the spatial autoregressive parameter
$\rho$:
\begin{equation*}
    G_{\rho i}=\frac{\partial P_i}{\partial \rho}=\frac{\partial}{\partial \rho}\left[1+exp\left(q_i\right)\right]^{-1}=\frac{\partial}{\partial \rho}\left[ 1+exp\left(\frac{-\iw^{-1}\xb _i'\bb}{\sigma_{e i}}\right) \right]^{-1},
\end{equation*}
where $\iw=(\Ib-\rho\Wb)$ and $\qb$ as in equation
(\ref{eq:stanadization}). We derive the following gradient
\begin{equation*}
    G_{\rho i} = P_i(1-P_i)\left(h_i+\frac{q_i}{\sigma_{e i}}\frac{\partial \sigma_{e i}}{\partial \rho}\right),
\end{equation*}
where ${\bm h}$ is defined in equation (\ref{eq:km_deriv}) and
\begin{align}
    \frac{\partial \sigma_{e i}}{\partial \rho} &= \frac{\partial}{\partial \rho} \left[(\iw'\iw)^{-1}\right]^{\frac{1}{2}}\nonumber\\
    &= -\frac{1}{2\sigma_{e i}}(\iw'\iw)^{-1}(\iw + \iw')\Wb(\iw'\iw)^{-1}\label{eq:km_corr_deriv}.
\end{align}
If $\iw$ is symmetric, equation (\ref{eq:km_corr_deriv}) simplifies to:
\begin{align*}
    \frac{\partial \sigma_{e i}}{\partial \rho} &=  -\frac{1}{\sigma_{e i}}\iw^{-1}\Wb\iw^{-1}\iw^{-1},
\end{align*}
which leads to equation (\ref{eq:km_deriv}).

\end{document}